\documentclass[a4paper,11pt]{article}

\usepackage[utf8]{inputenc}
\usepackage[T1]{fontenc}
\usepackage{lmodern}
\usepackage[english]{babel}
\usepackage{xspace}

\usepackage[normalem]{ulem}

\usepackage[a4paper,margin=0.94in]{geometry}

\usepackage{amsmath, amssymb, amsthm}

\usepackage{graphicx}
\usepackage{booktabs}
\usepackage{caption}
\usepackage{subcaption}
\usepackage{array}

\usepackage[colorlinks=true, allcolors=blue]{hyperref}
\usepackage{cite}

\newcommand{\appRef}[1]{app.~\ref{#1}\xspace}

\newcommand{\eqRef}[1]{eq.~\eqref{#1}\xspace}

\newcommand{\figRef}[1]{fig.~\ref{#1}\xspace}

\newcommand{\figsRef}[1]{figs.~\ref{#1}\xspace}

\newcommand{\secRef}[1]{sec.~\ref{#1}\xspace}

\newlength{\plotwidth}
\usepackage[dvipsnames]{xcolor}

\newcolumntype{F}{>{\color{Bittersweet}}c}
\newcolumntype{H}{>{\color{black}}c}
\newcolumntype{J}{>{\color{NavyBlue}}c}
\newcolumntype{S}{>{\color{black}}c}
\newcolumntype{R}{>{\color{Blue}}c}

\newcommand{\summaryTable}[1]{
\centering
{\bf #1}\\[0.5mm]
\begin{tabular}{|l|R|J|F|}
\hline  & 
~~R~~ & ~~J~~ & ~~F~~ \\
\hline
}

\newcommand{\mrm}[1]{\ensuremath{\mathrm{#1}}\xspace}

\title{\textbf{Cat's Cradle diagrams for substructure studies}}
\author{
    Peter Skands\thanks{School of Physics and Astronomy, Monash University, Wellington Road, Clayton, VIC-3800, Australia} \\
}
\date{September, 2026}

\begin{document}

\maketitle

\vspace*{3mm}\begin{center}
\includegraphics[width=0.52\textwidth]{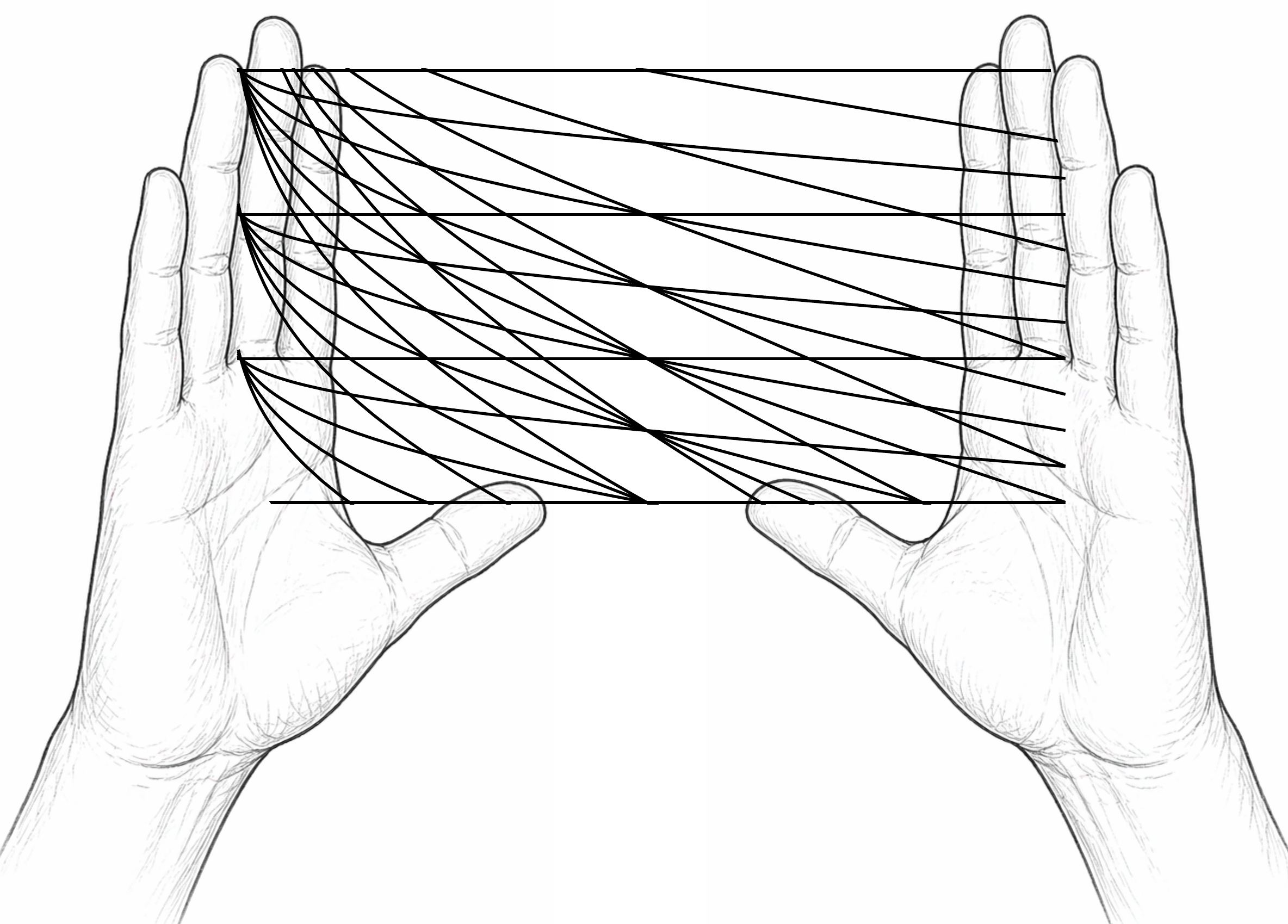}
\end{center}\vspace*{8mm}

\begin{abstract}
We propose ``Cat's Cradle diagrams'' as a framework to analyse the perturbative accuracy of gauge quantum field theories across different kinematic regions, interpolating between the power countings of fixed-order perturbation theory and soft-collinear resummation. We propose a simple intermediate counting aimed at quantifying parametric accuracy for resolved jet substructure, nested so that  $\mrm{N}^{n}\mrm{LO} \subset \mrm{N}^{n}\mrm{LJ} \subset \mrm{N^{n+1}LL}$. We show that the resolved jet-substructure power counting, $\mrm{N}^{n}\mrm{LJ}$, emphasises  contributions that are important for resolved substructure but which are lumped together with less dominant terms in both of the two other countings. It may
 therefore furnish a complementary quantification of relative parametric accuracy in a kinematic region ``midway'' between the hard (fixed-order) and soft-collinear (resummation) regimes.
We discuss practical applications to parton showers, in the context of a hard process with a characteristic hard scale of order 200 GeV. 
\end{abstract}
\vfill
\clearpage
\section{Introduction}
The perturbative series for a soft- and collinear-safe observable of a gauge quantum field theory can be cast as a two-parameter expansion in powers of the gauge coupling, $\alpha$, and logarithms of scale ratios, $L$, with generic coefficients proportional to
\begin{equation}
\alpha^n L^m\,,
\end{equation}
with $m \in [0, 2n]$. Specifically, for a generic physical process that involves a gauge-charged particle experiencing a momentum transfer characterised by a scale  $Q_H$, and an observable that is sensitive to radiation from that particle at a (lower) scale $Q_S$, the logarithms will be of the form
\begin{equation}
  L = \ln(Q_H/Q_S)\,.
\end{equation}
For problems with massive gauge-charged particles, ratios of $Q_S/m$ and/or $Q_H/m$ will also appear. And for SU(N) gauge theories, the parameter $1/N^2$ furnishes an additional power-counting parameter.

In this study, we focus on the simplest case of massless particles and, to keep the exposition uncluttered, let the $1/N_C^2$ counting for QCD be implicit via $\alpha_s^n/N_C^2 \sim \alpha_s^{n+1}$. 

Specifically, consider the cross section for events that have $Q_S$ smaller than some arbitrary (but still perturbative) $Q_\mrm{cut} \gtrsim {\cal O}(1~\mrm{GeV})$. For example, if $Q_S$ were a subjet resolution scale, then we would be talking about the integrated cross section for \emph{not} having any subjets resolved above $Q_\mrm{cut}$. Via unitarity, we can write this as the total inclusive cross section, $\Sigma_\mrm{inc}(Q_H)$, minus the integrated cross section for events that \emph{do} have resolved substructure above $Q_\mrm{cut}$:
\begin{eqnarray}
\Sigma(Q_S \le Q_\mrm{cut}) & ~=~ &  \Sigma_{\mrm{inc}}(Q_H) - \Sigma( Q_S \ge Q_\mrm{cut}) \\[3mm]
& = & \Sigma_{\mrm{inc}}(Q_H)\left( 1 - \frac{\Sigma( Q_S \ge Q_\mrm{cut})}{\Sigma_{\mrm{inc}}(Q_H)}
\right) \label{eq:radCor}
\end{eqnarray}
We will be interested in what can we say about the relative parametric accuracy that different perturbative approaches can offer (at best and at worst) for the second term inside the parenthesis in \eqRef{eq:radCor}, as a function of $Q_S/Q_H$.

For $Q_S \sim Q_H$,  $L$ is small and fixed-order truncations of perturbation theory are reliable, with each additional complete order corresponding to an improvement of the parametric uncertainty by a factor of order $\alpha_s$.  For $Q_S \ll Q_H$, however,  resummations of higher-order terms proportional to $\alpha_s^n L^m$ will be relevant. In this region, the main established accounting of systematic improvements is the one commonly used in all-orders soft-collinear resummation contexts. 
Here, we will focus on the large class of observables that admit exponentiation of the logarithmic terms, such that the full perturbative prediction can be cast in the form,
\begin{equation}
\Sigma \,=\, \Sigma_0(1 - \alpha - \alpha^2 - \ldots )\exp\left( -\sum_{n=1}^\infty\sum_{m=1}^{n+1} \alpha^n L^m\right)\,.\label{eq:exp}
\end{equation}
For such observables, including the $k$'th order of $\alpha_s^n L^{n+1-k}$ (for all $n$) yields a parametric improvement of order $1/L^k \sim \alpha_s^k$, called N$^k$LL accuracy. Note that \eqRef{eq:exp} should not be taken as literally expressing the form taken by such resummations but as representing an idealised full perturbative result; the $(1-\alpha-\alpha^2-\ldots)$ terms are intended to represent whatever is ``left'' at each complete order after separating out the terms that exponentiate and may still contain complicated kinematics dependence of their own, cf., e.g., \appRef{app:SudakovIntegrals}.

For parton showers (see, e.g.,~\cite{vanBeekveld:2026uxl}), the exponentiated terms, $\exp(-\ldots)$, are closely related to the Sudakov factor, but they are not identical. We emphasise that \eqRef{eq:exp} expresses the perturbative result \emph{after} showering, hence recoil effects in the shower may generate higher-order terms that are not directly present in the Sudakov factors themselves. E.g., the parton-shower Sudakov factor for the first emission will not account directly for the recoil that the first emission may receive from subsequent ones, while this effect \emph{would} need to be accounted for in \eqRef{eq:exp}. 

We shall use this type of counting (inside the exponent) throughout this work, mainly for simplicity, to demonstrate what we call cats-cradle diagrams in as simple a context as possible. Generalising this to the cross-section level will be the subject of a follow-up study. 

It is understood that each term in \eqRef{eq:exp} is associated with a coefficient which we shall assume can be taken to be of order unity. This is loosely motivated by the exact coefficients for the first-order double and single logarithms for gluon emission from a $q\bar{q}$ dipole, with $Q_S \sim p_T$, being $2C_F/\pi \sim 0.85$ and $3 C_F /\pi \sim 1.25$ respectively, and the coefficients generated via running-coupling effects likewise exhibiting less than a factor-2 deviation from unity per several orders in $\alpha_s$\footnote{If we define the generic logarithm to be $L\equiv \ln(p_\perp^2/s)$ we observe larger systematic deviations from the order-unity assumption, hence why we here define $L\equiv \ln(p_\perp/\sqrt{s})$, cf.~\appRef{app:SudakovIntegrals}.}. This is shown in \appRef{app:SudakovIntegrals}. To emphasise that any corresponding accuracy estimates are based solely on power countings, we will refer to them as ``parametric accuracies''.

Specifically, for this initial study we shall limit ourselves to thinking of $Q_S$ as a $p_T$-like resolution scale of a 3rd jet in a Born-level 2-jet event. I.e., $Q_S$ will represent the $p_T$ resolution scale of ``one-prong'' jet substructure. 
This choice of observable is as close as possible to the observable that $p_T$-ordered parton showers should resum by construction. Note, therefore, that in the context of parton showers we will really be discussing the maximum accuracy we expect them to be capable of achieving, for the (class of) observables they are constructed to resum. This is obviously a much lower bar than asking what their accuracy for any possible (IR safe\footnote{We do not distinguish between soft and collinear singularities in this work, lumping them both together as infrared (IR) singularities (as opposed to the UV ones that are related to renormalisation).}) observable would be, but nevertheless an important question which we would like to answer. One should then bear in mind that their effective accuracy for observables that are very different from ones exhibiting $p_T$-like logarithmic structures may be lower. 

For definiteness, we will use the \textsc{Ariadne} definition of $p_T$~\cite{Gustafson:1987rq,Lonnblad:1992tz,Giele:2007di},
\begin{equation}
p_{Tj}^2 ~=~\frac{s_{ij} s_{jk}}{s_{ijk}}~=~ E_j^2 (1-\cos\theta_{ij})(1-\cos\theta_{jk}) \, x_i \, x_k~\,
\end{equation}
which makes it simple to integrate at least the first-order tree-level matrix elements (done in  \appRef{app:SudakovIntegrals}) and also directly corresponds to the ordering variables used in \textsc{Ariadne}~\cite{Lonnblad:1992tz}, \textsc{Vincia}~\cite{Brooks:2020upa}, and \textsc{Apollo}~\cite{Preuss:2024vyu}. Is is also closely related to other $p_T$-ordered shower evolution variables. E.g., it becomes identical to Durham $k_T$ in the soft-collinear limit ($x_i\to 1,\, x_k\to1,\,\cos\theta_{jk}\to -1$) though it is always a bit smaller than Durham $k_T$ outside that limit. 

Eq.~\eqref{eq:exp} can be rewritten 
\begin{eqnarray}
\ln\left(\frac{\Sigma}{\Sigma_0}\right) & =&  \ln(1 - \alpha - \alpha^2 - \ldots) \, - \, \sum_{n=1}^\infty\sum_{m=1}^{n+1} \alpha^n L^m \\
& \approx & \,{- \alpha - \alpha^2 - \,\ldots\, } - \, \sum_{n=1}^\infty\sum_{m=1}^{n+1} \alpha^n L^m~, \label{eq:sum}
\end{eqnarray}
where the implicit constants multiplying each of the non-logarithmic terms are not identical between the first and second line --- but their parametric order is.
We shall use the latter form for comparing logarithmically enhanced and non-enhanced corrections on a similar footing.

\begin{figure}[t]
\centering
\includegraphics*[width=\plotwidth]{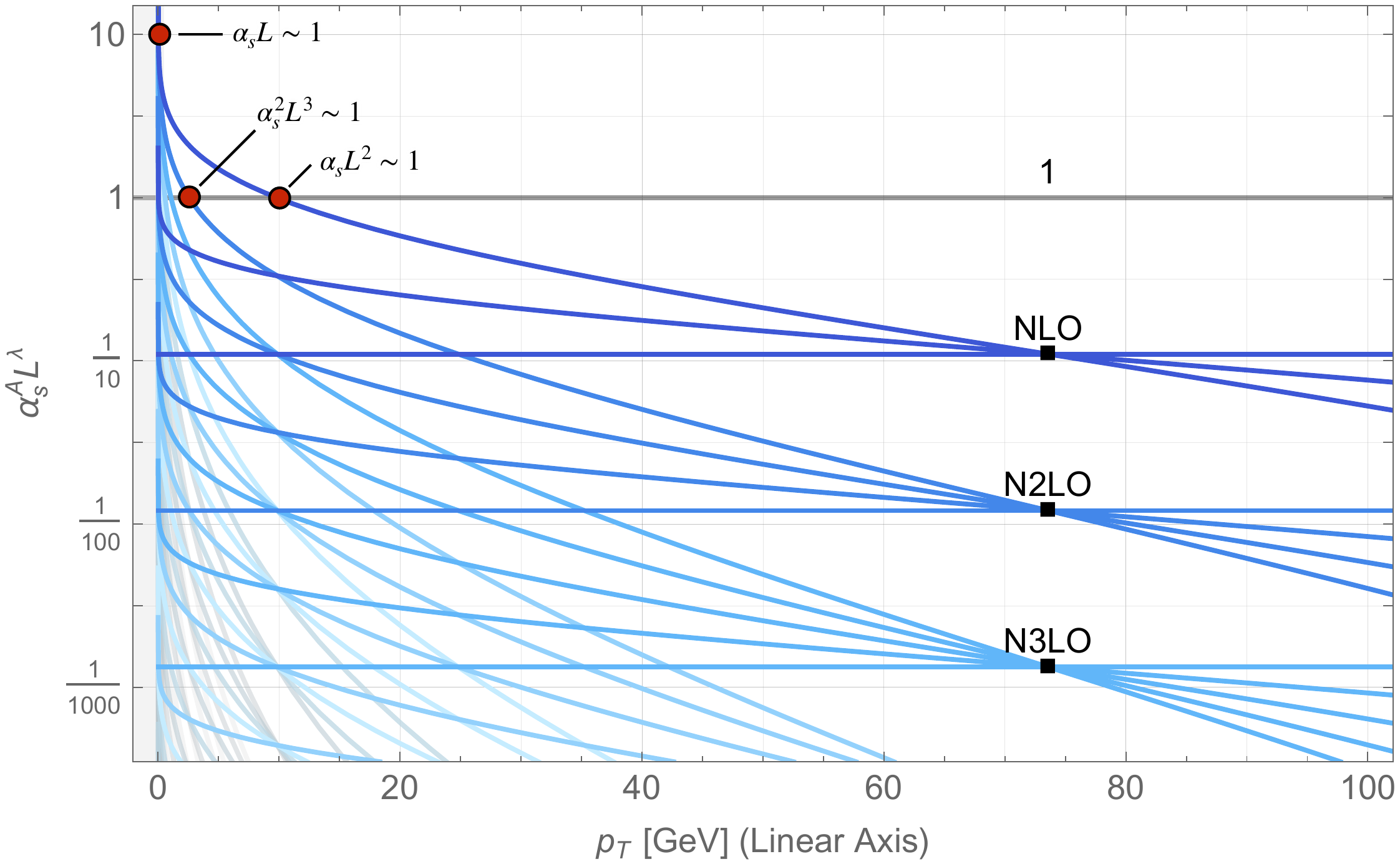}
\caption{Cats-cradle diagram illustrating the parametric sizes of generic perturbative coefficients, as a function of a $p_T$ substructure-resolution scale, for a pair of back-to-back hard jets with $M_{jj}\,=\,200~\mathrm{GeV}$. The colour coding corresponds to a ``fixed-order'' power counting: terms that share the same power of $\alpha_s$ (indicated by labels) are given the same colour.
\label{fig:generic}}
\end{figure}
In \figRef{fig:generic}, we show a generic version of what we call a ``cat's-cradle diagram'' (named for the children's game played with loops of string), illustrating the parametric size of each of the terms inside the exponent, for a representative hard system consisting of a colour-connected back-to-back quark-antiquark pair with a total CM energy of $\sqrt{s}~=~200$~GeV, and $\alpha_s(\sqrt{s}) \equiv 0.11$. 
The colour coding  corresponds to a ``fixed-order'' power counting: terms that share the same power of $\alpha_s$ (indicated by labels on the plot) are given the same colour.
Terms that would be included in an NLO calculation of the Born-level 2-jet process are shown in dark blue, NNLO would include also the terms shown in lighter blue, etc. 

We see that at a substructure scale of $p_T \sim 0.1 \frac{\sqrt{s}}{2} \sim 10\,\mrm{GeV}$ (where $\alpha_s L^2 \sim 1$), logarithmically enhanced higher-order terms reduce the parametric improvement to order $\sqrt{\alpha_s}$ for each additional full order of $\alpha_s$ included in the calculation.  Note that this value is for our chosen reference hard process of two 100-GeV back-to-back quark jets. For a jet pair with $\sqrt{s} = r \, (200\,\mrm{GeV})$, this resolution scale will likewise be greater or smaller by a factor $r$. 

In \secRef{sec:quant}, we define how we quantify relative orders of parametric improvement, as a function of $p_T/\sqrt{s}$, showing, e.g., the reduction described above from a full order $\alpha_s$ improvement for each fixed order at high scales $p_\perp \sim \sqrt{s}$, to a factor $\alpha_s^{1/2}$  at scales $p_\perp\sim 0.1\,\frac{\sqrt{s}}{2}$. We also contrast the fixed-order power counting to the soft-collinear resummation power counting, and illustrate the results in terms of colour-coded cats-cradle diagrams. This demonstrates that there is a wide region in between the fixed-order and resummation limits in which neither of these power countings offers an optimal quantification of the relative parametric accuracy. We call this region the resolved-substructure region, and propose a simple intermediate power counting targeting that region specifically.

In \secRef{sec:showers}, we bring the same style of analysis to bear on $p_T$-ordered dipole showers, and demonstrate that they tend to reach their maximum parametric accuracy in the resolved-substructure region, while their parametric accuracies tend to be lower in both the fixed-order and soft-collinear resummation regions. 
The former can be systematically improved by matching and merging~\cite{Catani:2001cc,Lonnblad:2001iq,Frixione:2002ik,Mangano:2006rw,Frixione:2007vw,Lopez-Villarejo:2011pwr,Lonnblad:2011xx,Cooper:2011gk,Frederix:2012ps,Lonnblad:2012ng,Lonnblad:2012ix,Alioli:2012fc,Hamilton:2012rf,Hartgring:2013jma,Hoche:2014uhw,Jadach:2015mza,Monni:2019whf,vanBeekveld:2025lpz}, the latter by the development of showers reaching higher logarithmic accuracies~\cite{Dasgupta:2020fwr,Forshaw:2020wrq,Nagy:2020rmk,Herren:2022jej,FerrarioRavasio:2023kyg,vanBeekveld:2024wws,Preuss:2024vyu,Helliwell:2025krf}. 
Nevertheless, our analysis demonstrates that, in the resolved-substructure region, even conventional ``LL'' dipole showers with local recoils are capable of reaching parametric accuracies that are about an order of magnitude better than pure LL resummations. We believe this conclusion was always implicitly understood in the community but hope that this analysis may shed some further light on it. (See also \cite{Hoche:2017kst} for important related work.)
For specific questions of resolved jet substructure, cats-cradle analysis may also help as a guide towards which further improvements are most important to pursue, to increase parametric accuracy in this specific region.

Finally, in \secRef{sec:conclusions}, we summarise and conclude.

%Note the ambiguities associated with renormalons and with inversion of Mellin transforms? Then ignore them. 

\section{Quantifying parametric accuracy for resolved jet substructure}
\label{sec:quant}

The information present in a full cats-cradle diagram like the one in \figRef{fig:generic} can be summarised compactly by defining a measure of relative parametric accuracy, 
\begin{equation}
R_\mathrm{acc}(p_T) ~=~\frac{\mbox{Largest Uncontrolled Term}\,(p_T)}{\mbox{Largest Controlled Term}\,(p_T)}~,
\end{equation}
with the numerator and denominator defined by the perturbative calculation whose relative parametric accuracy one wishes to quantify. Calculations that have small uncontrolled terms, relative to their controlled terms, have small values of $R_\mrm{acc}$. 

We define the corresponding effective ``order of parametric accuracy'' as
\begin{equation}
n_\mrm{eff}(p_T) ~= ~\log_{\alpha_s}\!\left(R_\mathrm{acc}(p_T)\right)~. \label{eq:neff}
\end{equation}
This simply converts $R_\mathrm{acc}$ into a corresponding effective number of powers of $\alpha_s$. Recall that $\alpha_s < 1$ and that 
\begin{equation}
\log_{\alpha_s} R_\mrm{acc} = \log_{1/\alpha_s} R_\mrm{acc}^{-1}~,
\end{equation}
hence calculations that have small $R_\mrm{acc} \le \alpha_s$ will have large $n_\mrm{eff} \ge 1$. Note that we shall here use $\alpha_s=\alpha_s(\sqrt{s})\equiv0.11$. 

\begin{figure}[t!p]
\centering
\includegraphics*[width=\plotwidth]{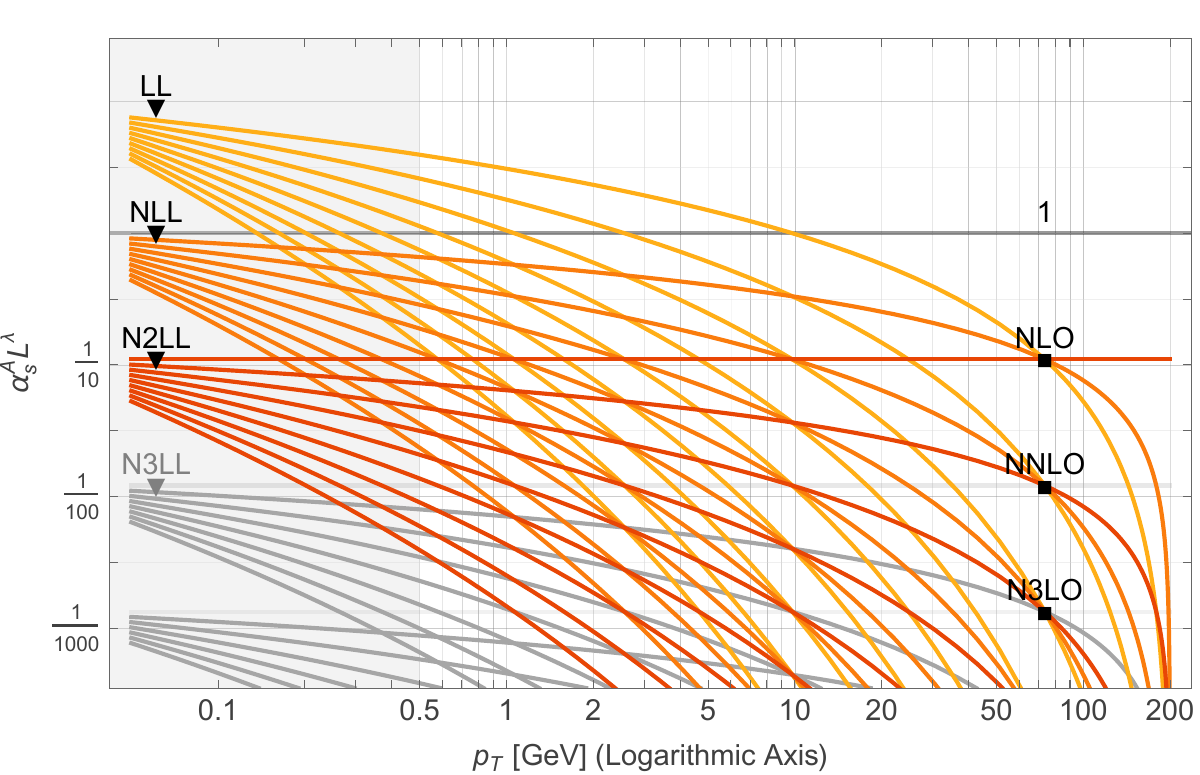}
\caption{Cats-cradle diagram on a logarithmic scale in the $p_T$ substructure resolution scale, colour coded according to a ``soft-collinear'' power counting: terms $\propto \alpha_s^A L^\lambda$ that share the same value of $A - \lambda$ (indicated by LL, NLL, etc.) are given the same colour.
\label{fig:resum}}
\end{figure}
In \figRef{fig:resum}, we show a cats-cradle diagram colour coded according to the soft-collinear power counting: terms $\alpha_s^A L^\lambda$ that share the same value of $A-\lambda$ are given the same colour, and are labelled LL, NLL, etc. To emphasise the logarithmic region, this plot is shown on a logarithmic scale in $p_T$, while the one in \figRef{fig:generic} used a linear scale.

Comparing \figsRef{fig:generic} and \ref{fig:resum}, we see that these two power countings simply correspond to different hierarchies among the perturbative terms but that both clearly represent systematic countings appropriate to their respective (hard or soft) kinematic domains. However, we also see that, while the hierarchy of terms is clear in the hard region (\figRef{fig:generic}) and in the deep infrared (\figRef{fig:resum}), it is not so towards the middle of the diagrams. There, terms that are formally subdominant in the soft-collinear region (e.g., N$^{k+1}$LL with respect to N$^{k}$LL) may be numerically larger than some or most of the previous-order ones. And likewise for terms that are subdominant in the fixed-order region (e.g., N$^{n+1}$LO with respect to N$^{n}$LO).

\begin{figure}[t]
\centering
\includegraphics*[width=\plotwidth]{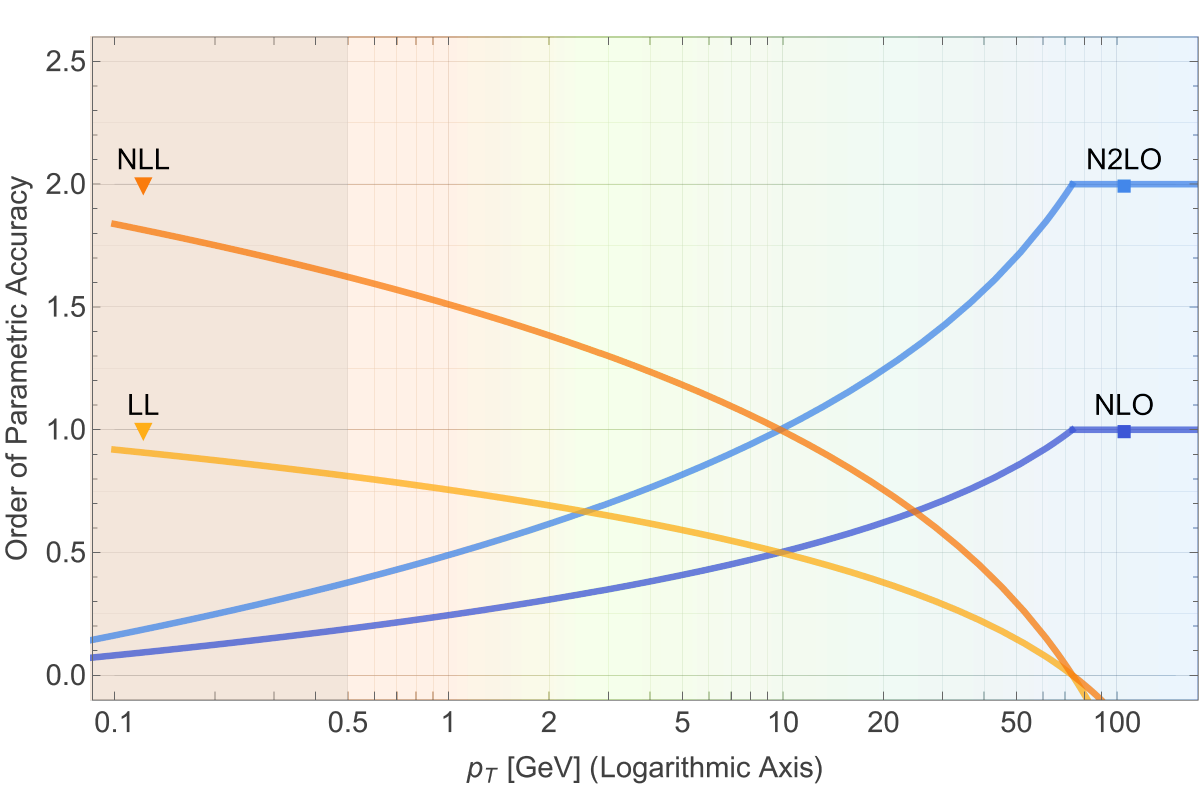}
\caption{Order of parametric accuracy, defined according to \eqRef{eq:neff}, as a function of \textsc{Ariadne} $p_T$, for LL, NLL, NLO, and N2LO calculations, with $\sqrt{s}\,=\,200~\mrm{GeV}$.
\label{fig:parAccResFO}}
\end{figure}

\begin{figure}[tp]
\centering
\includegraphics*[width=\plotwidth]{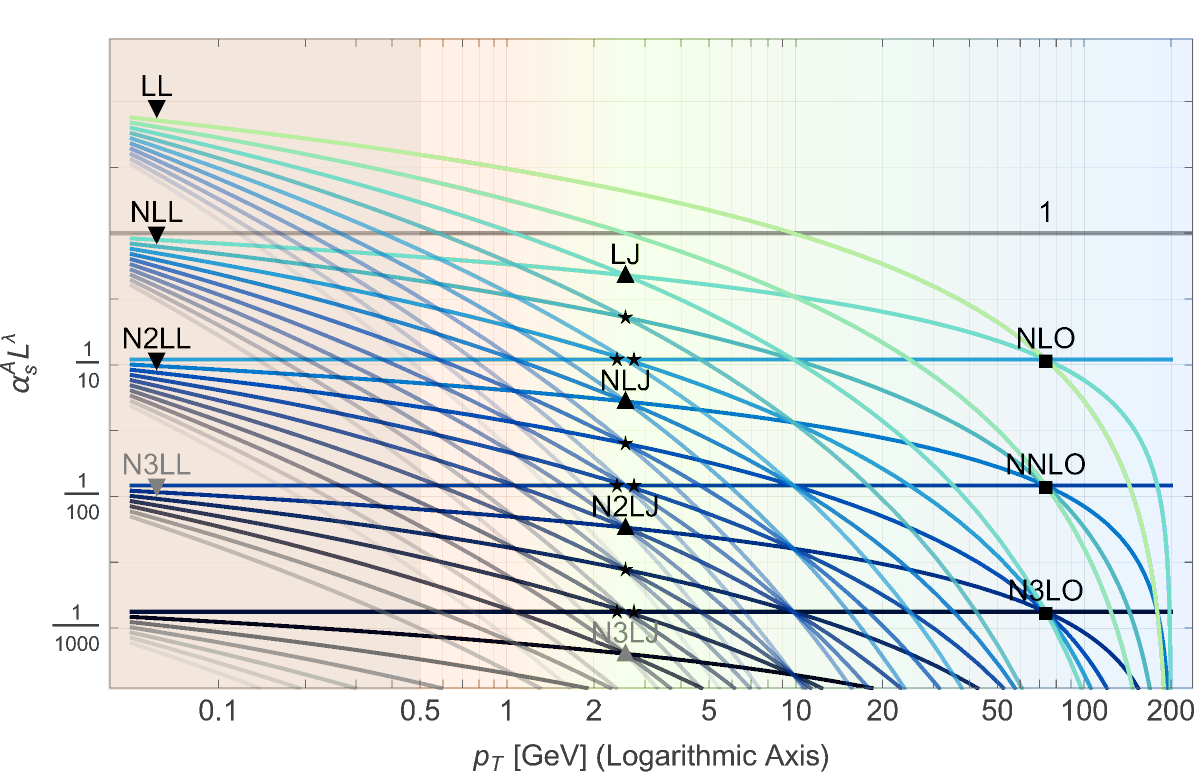}
\caption{Cat's-cradle diagram colour coded according to $\mrm{N}^n\mrm{LJ}$ counting, with each term sharing the same $n$ in  $\mrm{N}^n\mrm{LJ}$ assigned the same colour. \label{fig:jetcolours}}
\end{figure}

Thus, neither of the two power countings represented by the colour codings in \figsRef{fig:generic} and \ref{fig:resum}  provide an optimal systematic classification of the hierarchy of perturbative terms in the middle region. This can be quantified by plotting the corresponding effective orders of parametric accuracy, defined according to \eqRef{eq:neff}. This is shown in \figRef{fig:parAccResFO}, for NLO, N2LO, and LO + LL/NLL calculations, respectively.
We see that:
\begin{itemize}
\item Fixed-order calculations retain their full nominal parametric accuracy from the hard scale $\sqrt{s} = 200~\mrm{GeV}$ down to a scale of order $\sqrt{s}/e \sim 70~\mrm{GeV}$, below which logarithmically enhanced terms beyond the given fixed order begin to spoil their effective rate of convergence. As mentioned above, at $p_T\sim 10~$GeV the relative improvement is only half an effective order per full order of $\alpha_s$.
\item Soft-collinear resummations deliver a full order of improvement per logarithmic order in the far infrared, and continue to deliver almost a full order of improvement per log order up to about $p_T\sim 10~$GeV, at which point they likewise only deliver half an effective order of accuracy, per full log order.
\item Right in the middle, say for $5\,\mrm{GeV} \lesssim p_T \lesssim 20\,\mrm{GeV}$, there is the region we call the ``resolved jet substructure region'', in which the effective convergence rates of both types of power countings are comparably slow.  
\end{itemize}
We therefore propose to add a complementary power counting designed so that each full order in this counting corresponds to a full order of parametric improvement in the resolved jet-substructure region. There is of course an infinite family of different choices that could be made, one of which would correspond to so-called ``double-log'' counting\footnote{Note, however, that double-log counting is normally done at the cross-section level rather than in the exponent.}, anchored at the point $\alpha_s L^2 = 1$. An alternative simple one that appears to us to have some useful properties is to assign all terms $\alpha_s^A L^\lambda$ that share the same value of $3A-2\lambda$ the same order of resolved jet substructure. A cat's-cradle diagram with this colour coding is shown in \figRef{fig:jetcolours}. 
Specifically, a term proportional to $\alpha_s^A L^\lambda$ is classified as:
\begin{center}
\begin{tabular}{lcl}
 Fixed Order
 & $\mrm{N}^n\mrm{LO}$ & with $n=A$\\ 
Jet-Substructure Order  & $\mrm{N}^n\mrm{LJ}$ & with $n = A + 2(1-\lambda)/3 - 1$\\
Soft-Collinear Order 
& $\mrm{N}^n\mrm{LL}$ & with $n = A + (1-\lambda)$
\end{tabular}
\end{center}

One sees immediately that lines with different colours intersect at both the left (soft-collinear) and right (fixed-order) ends of the plot range, hence this counting is not systematic in either of those two extremes, while (by construction) there is now an ordered gradient in a region around the LJ $\to$ NLJ $\to$ ... vertical, where this counting can therefore be regarded as systematic. 

\begin{figure}[tp]
\centering
\includegraphics*[width=\plotwidth]{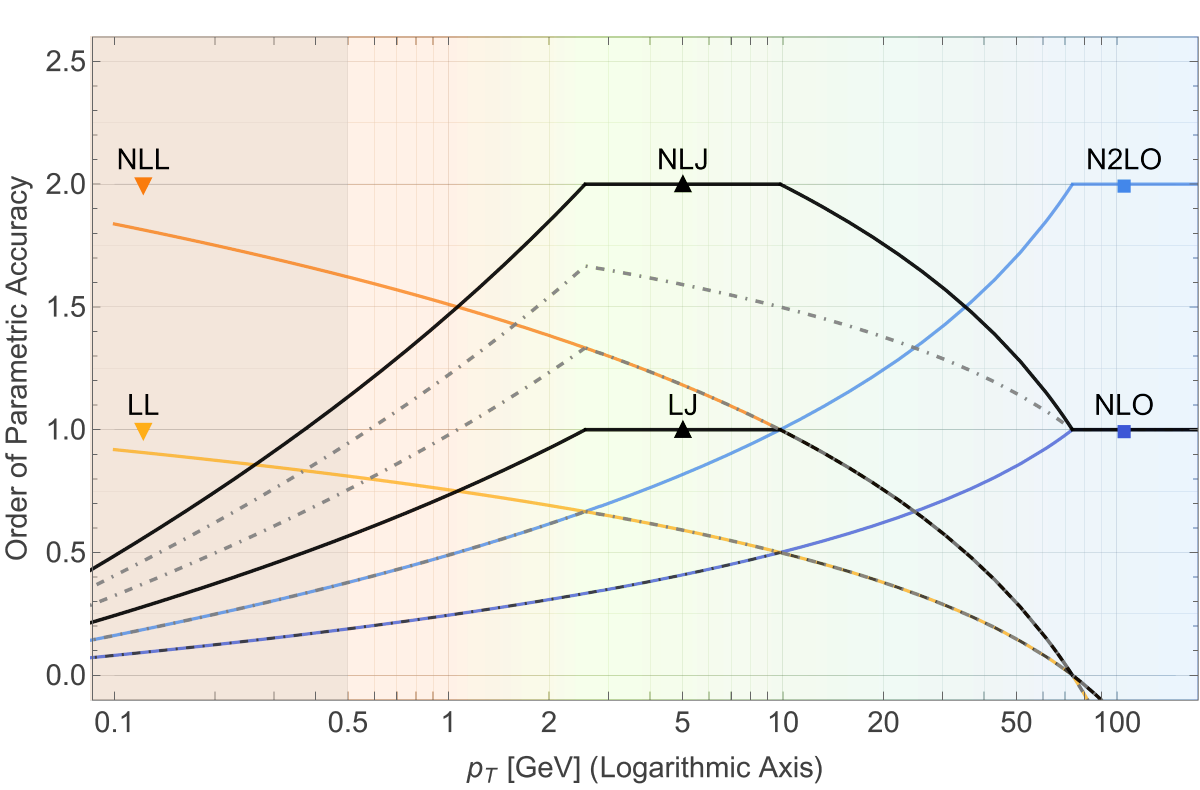}
\caption{Order of parametric accuracy, defined according to \eqRef{eq:neff}, as a function of \textsc{Ariadne} $p_T$, for $\sqrt{s}\,=\,200~\mrm{GeV}$ and $\alpha_s(\sqrt{s})\equiv 0.11$. Fractional J orders are shown with dot-dashed lines.\label{fig:neffJ}}
\end{figure}

One also sees that, due to the density of lines in this intermediate region, fractional ``J'' orders are possible. With our choice of counting, the ``J'' order increments by 1/3 for each additional (set of) terms that are included, represented by small $\star$ symbols in \figRef{fig:jetcolours}. Here, however, we shall focus on the integer orders, LJ, NLJ, etc., which are the orders that correspond to integer values of $n_\mrm{eff}$ in \eqRef{eq:neff}. 
This is illustrated in \figRef{fig:neffJ}, 
with gray dot-dashed lines indicating the fractional ``J'' orders. Incidentally, the lowest dot-dashed line corresponds to the so-called double-log approximation (DLA).

The terms that each ``J'' order includes in the exponent, are:
\begin{center}
\begin{tabular}{ll}
DLA & $\alpha_s L^2$ \\[1mm]
DLA$^\star$ & ~+~ $\alpha_s^2 L^3$ \\[1mm]
\textbf{LJ}  & ~+~ $\alpha_s^3 L^4$, $\alpha_s L$,\\[1mm]
LJ$^\star$  & ~+~ $\alpha_s^4 L^5$, $\alpha^2_s L^2$ \\[1mm]
LJ$^{\star\star}$  & ~+~  $\alpha_s^5 L^6$, $\alpha_s^3 L^3$, $\alpha_s L^0$
\\[1mm]
\textbf{NLJ}  & ~+~  $\alpha_s^6 L^7$, $\alpha_s^4 L^4$, $\alpha^2_s L^1$
\\[1mm]
NLJ$^\star$  & ~+~  $\alpha_s^7 L^8$, $\alpha_s^5 L^5$, $\alpha^3_s L^2$
\\[1mm]
NLJ$^{\star\star}$  & ~+~  $\alpha_s^8 L^9$, $\alpha_s^6 L^6$, $\alpha^4_s L^3$, $\alpha^2_s L^0$
\\[1mm]
\textbf{NNLJ}  & ~+~  $\alpha_s^9 L^{10}$, $\alpha_s^7 L^7$, $\alpha^5_s L^4$, $\alpha^3_s L^1$
\end{tabular}
\end{center}
Notice that at each integer J order (highlighted in \textbf{bold}), all singular terms through a specific fixed order are exponentiated: at LJ, all singularities at ${\cal O}(\alpha_s)$ are exponentiated; at NLJ, all singularities through ${\cal O}(\alpha_s^2)$ are exponentiated; etc.

Note also that, as we have defined it here, the ``J'' orders never include a full infinite-order sum of terms inside the exponent. This forces their order of parametric accuracy to still go to zero at the left-hand edge of the plots. If desired, one could modify the ``J'' counting, e.g., so that NLJ would subsume all terms that are included at LL. This would make the NLJ line asymptote to the LL one for small $Q_S$. We have elected not to do so here, as one may as well state all three parametric accuracies, to give a more detailed picture over the full domain of kinematics. I.e, for a given perturbative calculation, we would give its fixed-order accuracy (for its expected parametric accuracy in the hard region), its ``J'' order (for the resolved jet-substructure region), and its log-resummation order (for the soft-collinear region). 

We note, however, that N$^{n+1}$LL subsumes all N$^{n}$LJ terms (and all N$^{n}$LO ones). Hence a calculation that is demonstrated to have N$^{n+1}$LL accuracy will always have at least N$^{n}$LJ and N$^{n}$LO accuracy as well. The converse, however, is not true. A calculation can have a high fixed- and/or ``J''-order accuracy (as defined here) without having any specific soft-collinear accuracy. That is, e.g., the case for conventional parton showers, whose parametric accuracy for resolved substructure tends to be higher than their purely soft-collinear accuracy indicates. This is the topic of the next section.

\section{Parton showers \label{sec:showers}}

\begin{figure}[tp]
\centering
\includegraphics*[width=\plotwidth]{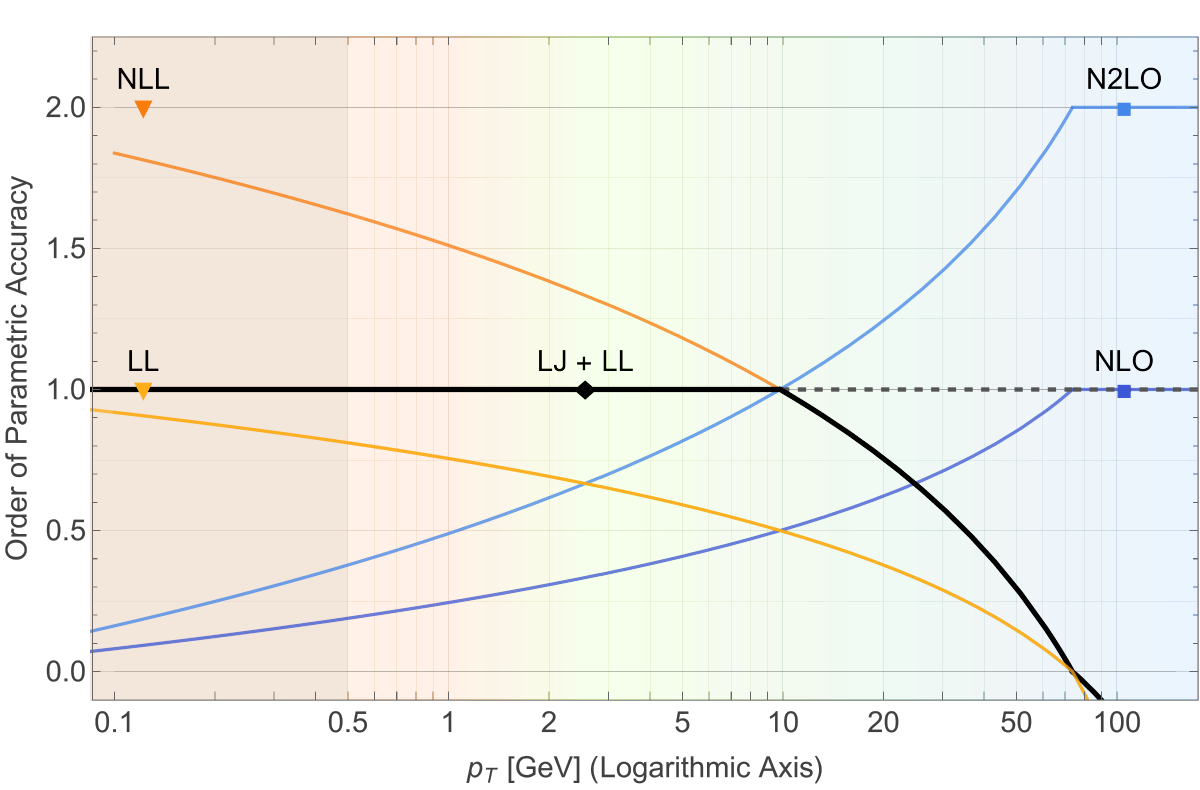}
\caption{Order of parametric accuracy,  \eqRef{eq:neff}, as a function of \textsc{Ariadne} $p_T$, for $\sqrt{s}\,=\,200~\mrm{GeV}$ and $\alpha_s(\sqrt{s})\equiv 0.11$.\label{fig:LJplusLL}}
\end{figure}

Consider a generic $p_T$-ordered parton shower, with a running $\alpha_s$ evaluated at a renormalisation scale $\sim p_T$, whose splitting kernels reproduce the first-order DGLAP kernels in collinear limits and the first-order leading-colour eikonal factors in soft limits. 
This would be true, e.g., of most (if not all) dipole/antenna showers. For a $p_T$-like one-prong substructure observable, without trimming or grooming, whose perturbative expansion starts at ${\cal O}(\alpha_sL^2)$, any such shower would include the correct first-order logarithmic terms in the exponent\footnote{Recall that we here lump subleading colour together with higher powers of $\alpha_s$ through $\alpha_s^n/N_C^2 \sim \alpha_s^{n+1}$.},
\begin{equation}
\alpha_s L^2 \,+\, \alpha_s L~.
\label{eq:PS}
\end{equation}
Already at this ``first-order'' sandbox level, we see that parton showers, from birth, are neither pure LL nor pure NLL calculations, since they 
intrinsically mix terms that, in the soft-collinear power counting, are classified as belonging to different orders, here LL ($\alpha_sL^2$) and NLL ($\alpha_s L$). From the point of view of LL accuracy, the second term in \eqRef{eq:PS} could be dropped entirely. However, in the resolved-substructure counting proposed in the preceding section, it is precisely this term that makes the difference between a calculation that only achieves LL accuracy and one that achieves LJ + LL accuracy. This allows us to quantify in what sense our sandbox parton shower has a higher accuracy than a corresponding purely LL soft-collinear resummation: it includes a coefficient which is irrelevant (subleading) in the soft-collinear region but which becomes relevant in the resolved-substructure region. This is illustrated in \figRef{fig:LJplusLL},
which also highlights another desirable feature of calculations that achieve simultaneous LJ + LL accuracy: there is a broad region of constant effective parametric accuracy, spanning the full range from the soft-collinear region to the hard substructure region. The quality of the approximation (as represented by the $n_\mrm{eff}$ measure) remains the same over this whole range. 

The dashed extension to the LJ + LL line in \figRef{fig:LJplusLL} illustrates the effect of including a first-order matrix-element correction, to achieve NLO + LJ + LL accuracy, in which case one has a single order of parametric accuracy throughout the entire domain of the observable.\footnote{Note that the second term in \eqRef{eq:PS} is exponentiated in an NLO + LJ + LL calculation. In a pure NLO + LL calculation, there is in principle an ambiguity on how to treat this term since it is divergent but not part of the pure LL resummation.}  

As highlighted in \cite{Hoche:2017kst}, parton showers also include several other effects:
\begin{enumerate}
\item Exact energy and momentum conservation. This is achieved via  kinematic mappings (a.k.a., recoil strategies) that can either be local (e.g., iterated $2\mapsto 3$ kinematic maps, as in local dipole/antenna showers~\cite{Gustafson:1987rq,Lonnblad:1992tz,Sjostrand:2004ef,Giele:2007di,Dinsdale:2007mf,Platzer:2009jq,Lopez-Villarejo:2011pwr,Hoche:2015sya,Brooks:2020upa}) or global (i.e., involving one or more particles outside the emitting colour dipole, as in \textsc{Herwig}'s angular-ordered showers~\cite{Marchesini:1983bm,Gieseke:2003rz}, \textsc{Deductor}~\cite{Nagy:2007ty},  \textsc{PanGlobal}~\cite{Dasgupta:2020fwr,Hamilton:2020rcu}, \textsc{Alaric}~\cite{Herren:2022jej},  \textsc{Apollo}~\cite{Preuss:2024vyu}, and jet-recoil schemes~\cite{Helliwell:2025krf}). 
\item Exact unitarity:
\begin{equation}
\mbox{Virtuals}~=~-\int \mbox{Reals}~,
\end{equation}
enforced by detailed balance in the shower evolution, cf., e.g.,~\cite{Hoche:2017kst,Altmann:2025yip}.
\item Use of the so-called CMW scheme for the running strong coupling~\cite{Catani:1990rr}.
\end{enumerate}
In the context of $p_T$-ordered showers, iterated $2\mapsto 3$ kinematics maps (i.e., local recoils) are known to generate undesired terms starting from the $\alpha_s^2L^2$ level~\cite{Dasgupta:2018nvj} and hence such showers cannot achieve full NLL accuracy. Global recoil strategies, however, make it possible to preserve the correct $\alpha_s^2L^2$ coefficient and hence retain NLL accuracy~\cite{Dasgupta:2018nvj,Dasgupta:2020fwr}. 

As mentioned above, there are by now several parton-shower algorithms that do possess NLL accuracy and even ones that have been demonstrated to reach NNLL accuracy~\cite{FerrarioRavasio:2023kyg,vanBeekveld:2024wws}. Nevertheless, showers that are only formally LL accurate, such as the current default ones in \textsc{Pythia}~\cite{Sjostrand:2004ef,Bierlich:2022pfr} and \textsc{Vincia}~\cite{Brooks:2020upa},  
remain in wide use. We therefore consider it interesting to examine the parametric accuracy of these showers in the region of resolved jet substructure.

\begin{figure}[tp]
\centering
\includegraphics*[width=0.95\textwidth]{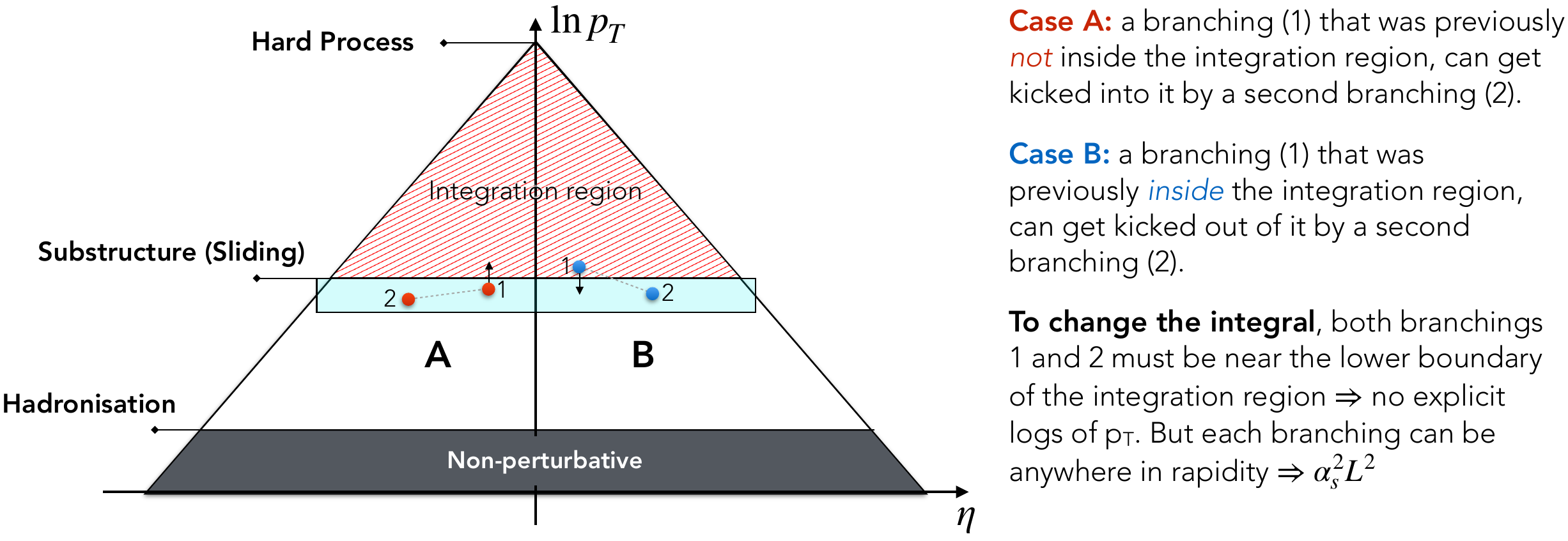}
\caption{Illustration of two cases in which recoil effects from higher-order corrections generate modifications to an integrated jet cross section.
\label{fig:plotRecoils}
}
\end{figure}

A simple illustration of the problem that local recoil strategies generate is given in \figRef{fig:plotRecoils}.
The top part of the phase space, shaded red, represents a region over which we integrate a jet cross section, down to a given $\ln(p_T)$ substructure scale. Below this region are illustrated two cases for which a second branching, occurring below the substructure scale, can change the value of the integral above it. In case A, illustrated on the left-hand side of the diagram, a first branching (1) was originally just below the integration region and hence would not have contributed to the integrated jet rate at first order. However, due to the recoil from a second branching (2), parton (1) can be kicked into the integration region. Case B, illustrated on the right-hand side of the diagram, shows the converse, a branching (1) that was originally inside the integration region and hence contributed to the integral at first order, can be kicked out of the integration region by the recoil from a second branching (2). The problem with local momentum mappings is that the second branching can still have this effect even if it is very far from the first in rapidity; this violates the factorisation of the corresponding double-emission matrix element for widely separated emissions. 

Determining the value of the coefficient of the violation is not currently a simple task. For our discussion, we note merely that averaging over azimuthal angles induces a partial compensation between the A and B mechanisms in \figRef{fig:plotRecoils}. A random walk in $p_T$ implies that the probability for A is 2/3 while the probability for B is 1/3. For angular-averaged one-prong observables, we therefore consider it reasonable to reduce the parametric $\alpha_s^2L^2$ violation by a factor $\sim1/3$. This also seems to be consistent with the findings of~\cite{Dasgupta:2018nvj} for the Cambridge algorithm and for the $\mathrm{FC}_1$ moment of the energy-energy correlation defined in~\cite{Banfi:2004yd}.

\begin{figure}[tp]
\centering
\includegraphics*[width=\plotwidth]{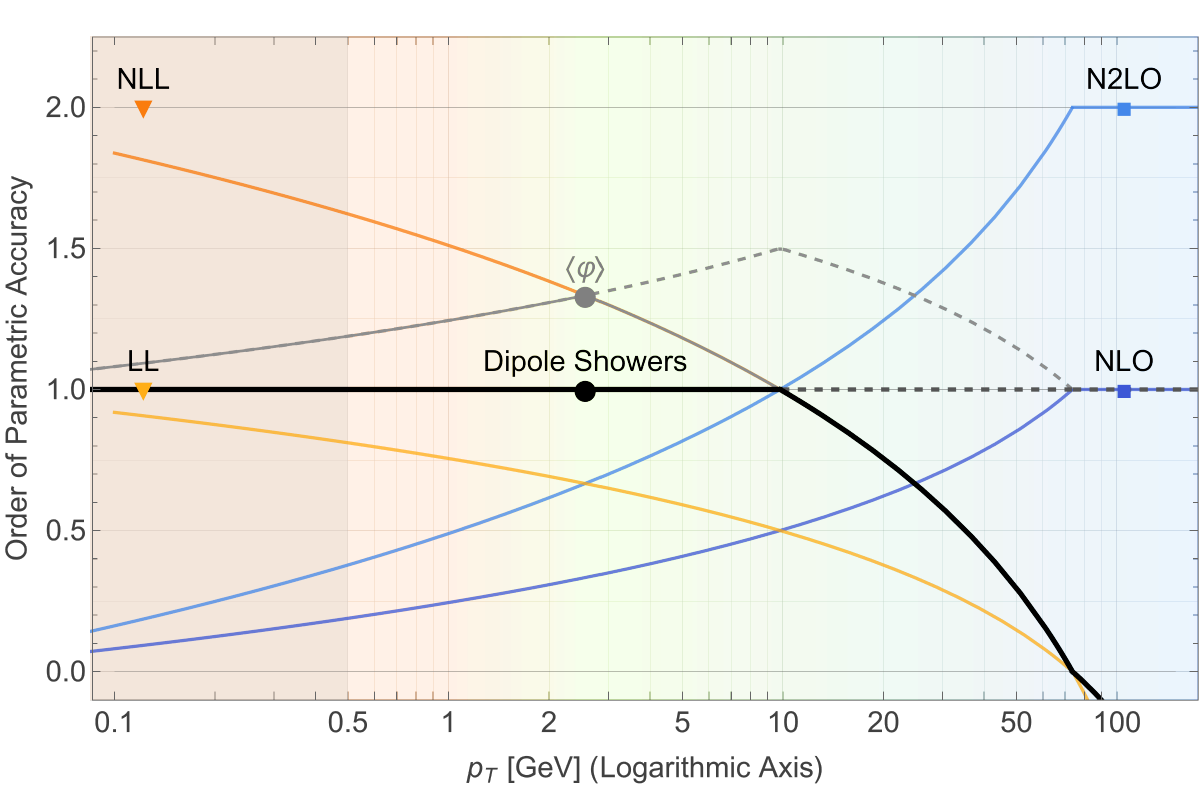}
\caption{Order of parametric accuracy,  \eqRef{eq:neff}, as a function of \textsc{Ariadne} $p_T$, for $\sqrt{s}\,=\,200~\mrm{GeV}$ and $\alpha_s(\sqrt{s})\equiv 0.11$, for dipole showers with and without assuming azimuthal averaging, compared to NLO, NNLO, and LO + LL/NLL.\label{fig:dipoleShowers}}
\end{figure}

The general parametric accuracy we expect of dipole showers (with LL recoils) is shown in \figRef{fig:dipoleShowers}, together with the accuracy we expect for azimuthally averaged quantities, labelled $\left<\varphi\right>$, which is higher due to the factor 1/3 reduction. 
The dashed lines show the further improvement if first-order matrix-element corrections are also included.

Despite our analysis admittedly being simple-minded, we think it is worth highlighting that we are led to the conclusion that the parametric accuracy of conventional dipole showers remains significantly higher than that of pure LL resummations across the full region of resolved substructure. And, for azimuthally averaged observables, dipole showers appear to remain consistent with NLL resummation down to fairly low scales of order a few GeV, for our reference case of back-to-back quark jets with $\sqrt{s}=200\,\mathrm{GeV}$. (For larger hard scales and/or gluon colour factors, all of the substructure scales move up correspondingly.)

This obviously still depends on the order-unity assumption we made for our generic perturbative coefficients. This would be important to test at higher orders, and would be a crucial step towards putting this kind of parametric study on a firmer footing, as would going to the cross-section (non-exponentiated) level, more general observables, and including further known types of coefficients such as power corrections, threshold logarithms, powers of $\alpha_s\ln^2(-1) = -\alpha_s\pi^2$, etc. 

\section{Conclusion}
\label{sec:conclusions}
This work began with the trivial observation that the perturbative coefficients generated by conventional parton-shower algorithms (in particular, dipole-style ones) include more than just the terms that are included in corresponding soft-collinear resummation limits. At lowest exponentiated order, 
\begin{eqnarray}
\mbox{LL Resummation (LLR)}&~:~&\alpha_s L^2 \,+\,\mbox{running coupling}\,,\\
\mbox{Parton-Shower Resummation (PSR)}&~:~&\alpha_s L^2\,+\,\alpha_sL\,+\,\mbox{running coupling}\,.
\end{eqnarray}
The $\alpha_s L$ term is formally subleading in the soft-collinear limit ($\alpha_s L \sim 1$) but can be numerically relevant at scales intermediate between the soft-collinear and hard-process scales, in what we call the region of resolved jet substructure. 

Based on simple parametric arguments, we illustrated in \figRef{fig:parAccResFO} that the region of a resolved 3rd jet in a Born-level 2-jet event is characterised by slower-than-parametric convergence for both soft-collinear and fixed-order power countings, with only approximately half an order of parametric improvement gained for each full order in the respective power countings. 

We propose a simple intermediate power counting, N$^n$LJ, according to which each full integer order corresponds to a full order of parametric accuracy gain in this region. 

We then demonstrated that the terms generated by parton-shower algorithms have a natural correspondence with this kind of power counting, and that conventional dipole-style showers can be classified as being LJ+LL accurate, rather than just LL. In \figRef{fig:dipoleShowers}, we illustrated that this corresponds to a very broad region of stable parametric accuracy, which also extends into the hard region if hard matrix-element-corrections are included in the description. We also made an argument for azimuthally-averaged observables having a greater parametric accuracy than non-averaged ones, staying numerically close to NLL parametric accuracy down to fairly low resolution scales, below which they then transit back to LL accuracy. 

To illustrate these points in as simple a context as possible, we took many shortcuts: 1) we limited our discussion to one quite specific type of observable, designed to be as close as possible to the ordering variable of  $p_\perp$-ordered shower algorithms; 2) we only considered terms of the form $\alpha_s^AL^\lambda$ inside the exponent, and only considered fixed-order corrections by recasting them on the same form, though we do show the exact integrals of the first-order matrix element in \appRef{app:SudakovIntegrals}; 3) we only tested our order-unity assumption for the perturbative coefficients at lowest order though we did allow for running-coupling effects; 4) we lumped subleading-colour effects together with higher-order effects via $1/N_C^2\sim \alpha_s$; 5) we only considered ``one-prong'' jet substructure; 6) we only considered pure final-state emissions from a fixed $q\bar{q}$ dipole. 

We do not claim that any of the results presented here are genuinely new. Indeed, many of the same points were already made in \cite{Hoche:2017kst}. The cats-cradle diagrams and the proposed N$^n$LJ power counting are new, though the latter is really a very close cousin of the already existing double-log counting; the counting used here just exhibits a wider plateau of stable parametric accuracy, cf.~\figRef{fig:neffJ}. Our main intent is to call attention to the phenomenological relevance of the region of resolved jet substructure, for which cats-cradle diagrams and some form of related intermediate power counting --- between the soft-collinear resummation and fixed-order ones --- may be helpful to analyse and quantify orders of parametric perturbative accuracy. We believe this may be particularly relevant for parton-shower algorithms, which suffer from being ``neither fish nor fowl'' according to the conventional established power countings, and have attempted to demonstrate this within a comparatively simple context.

\section*{Acknowledgements}
Many thanks to Jack Helliwell and Ludovic Scyboz for discussions and comments on this manuscript which helped to improve its rigour. Thanks also to Lenny McNaughton of Aspendale Primary School for bringing the game of cat's cradle to my attention.
This work was supported by
the Australian Research Council under Discovery Project grant DP220103512.

\appendix
\section{First-Order Sudakov Integrals and Running-$\mathbf{\alpha_s}$ Coefficients}
\label{app:SudakovIntegrals}
The spin- and colour-summed first-order squared matrix element for $Z\to q_ig_j\bar{q}_k$ (with massless quarks) is~\cite{Ellis:1980wv,Gehrmann-DeRidder:2005btv}:
\begin{equation}
|M_{3}|^2~=~\frac{2C_F\, g_s^2}{M_Z^2}\, \left(
\frac{2 y_{ik}}{y_{ij}y_{jk}} \,+\,\frac{y_{ij}}{y_{jk}} \,+\,
\frac{y_{jk}}{y_{ij}} 
\right)|M_{2}|^2~,
\end{equation}
with dimensionless invariants $y_{ab} \equiv 2p_a\cdot p_b$, $g_s^2 = 4\pi\alpha_s$, and the Born-level matrix element, $|M_{2}|^2$, only depending on the total invariant mass, $s = m_Z^2$ (in addition to the Born-level $Z\to q\bar{q}$ couplings).

The invariant one-particle antenna phase space is
\begin{equation}
\mrm{d}\Phi_{2\mapsto 3} ~=~\frac{m_Z^2}{16\pi^2 }\,\mrm{d} y_{ij} \,\mrm{d}y_{jk}\,\frac{\mrm{d}\phi_j}{2\pi}~.
\end{equation}

Defining invariant $p_\perp$ as in \textsc{Ariadne} and \textsc{Vincia}, 
\begin{eqnarray}
x_{\perp j}^2 & ~=~ & y_{ij} \,y_{jk}~,\\[2mm]
p_{\perp j}^2 & = & x_{\perp j}^2 \,m_Z^2
\end{eqnarray}
and invariant rapidity as
\begin{equation}
\eta_j ~=~\frac12 \ln\left(\frac{y_{jk}}{y_{ij}}\right)~,
\end{equation}
with inverse relations
\begin{eqnarray}
y_{ij}~=~x_{\perp j}\,e^{-y_j} & ~~,~~ & y_{jk}~=~x_{\perp j}\,e^{y_j}~,
\end{eqnarray}
the Jacobian for translating from $(y_{ij},y_{jk})$ to $(x_{\perp j}^2,\eta_j)$ is unity, hence 
\begin{equation}
\mrm{d}\Phi_{2\mapsto 3}~=~ \frac{m_Z^2}{16\pi^2} \, \mrm{d}x_{\perp j}^2\, \mrm{d}{\eta_j}\,\frac{\mrm{d}\phi_j}{2\pi}~.
\end{equation}
Expressed in the same variables, the ratio of the 3-particle to the 2-particle matrix element is:
\begin{equation}
\frac{|M_3|^2}{|M_2|^2}\,\mrm{d}\Phi_{2\mapsto 3}~=~\frac{C_F \alpha_s}{2\pi}\left[\,\frac{2}{x_{\perp j}^2} \,-\,\frac{4 \cosh(\eta_j)}{x_{\perp j}}\,+\,2\cosh(2\eta_j)\,\right]\,\mathrm{d}x_\perp^2\,\mrm{d}\eta_j\,\frac{\mrm{d}\phi}{2\pi}~.\label{eq:antExact}
\end{equation}
The kinematically accessible rapidity range  (without any approximation), as a function of $x_{\perp j}$, is:
\begin{equation}
\Delta \eta_j(p_\perp^2) ~=~\ln\left(
\frac{1+\sqrt{1-4x_{\perp j}^2}}{1-\sqrt{1-4x_{\perp j}^2}}
\right)~,\label{eq:rapRange}
\end{equation}
which is zero at $x^2_{\perp j}=1/4$, above which the physical phase space is closed. 
Defining
\begin{equation}
r_\perp = \frac12\left(1-\sqrt{1-4x_\perp^2}\right) ~~~\longleftrightarrow~~~x_\perp^2~=~r_\perp(1-r_\perp)~,
\end{equation}
the integrals of the three terms in \eqRef{eq:antExact}, integrated over the exact rapidity range, \eqRef{eq:rapRange}, from a fixed $r_\perp$ to its maximum value of 1/2 are:
\begin{align}
\displaystyle \int\frac{2}{x_{\perp j}^2} \,\mathrm{d}\Phi_{2\mapsto 3}& ~=~ 
\ln^2(r_\perp)\,+\,2\ln(r_\perp)\ln(1-r_\perp)\,-\,\ln^2(1-r_\perp)\,-\,\frac{\pi^2}{3}\,+4\mrm{Li}_2(r_\perp)\,
,\\[3mm]
\displaystyle\int \frac{-4\cosh(\eta_j)}{x_{\perp j}} \,\mathrm{d}\Phi_{2\mapsto 3}& ~=~4\ln(r_\perp) \,-\,4\ln(1-r_\perp)\,-\,16r_\perp \,+\,8\,,\\[3mm]
\displaystyle\int 2\cosh(2\eta_j)\,\mrm{d}\Phi_{2\mapsto3} &~=~
-\ln(r_\perp) \,+\,\ln(1-r_\perp)\,+\,4r_\perp -2\,,
\end{align}
with infrared ($r_\perp\to 0 ~\longleftrightarrow~ x_{\perp j}^2\to 0$ ) limits,
\begin{align}
& 
 \ln^2(r_\perp) 
 \,-\,\frac{\pi^2}{3}
 \,-\,2r_\perp\ln(r_\perp)
 \,+\,4r_\perp
&,~~~&
 \ln^2(x_{\perp j}^2)
 \,-\,\frac{\pi^2}{3}
 \,+\,4x_{\perp j}^2 
\,,\\[3mm]
& 
 4\ln(r_\perp) 
 \,+\,8 
 \,-\,12r_\perp 
&,~~~ & 
 4\ln(x_{\perp j}^2) 
 \,+\, 8
 \,-\, 8x_{\perp j}^2 
\,,\\[3mm]
& 
 -\ln(r_\perp) 
 \,-2\,
 \,+\, 3r_\perp 
&,~~~ & 
 -\ln(x_{\perp j} ^2) 
 \,-\,2
 \,+\,2x_{\perp j}^2 
\,,
\end{align}
respectively, combining to
\begin{equation}
\frac{C_F \alpha_s}{2\pi}\left( 
 \ln^2(x_{\perp j}^2)
 \,+\,3\ln(x_{\perp j}^2) 
 \,+\,6
 \,-\,\frac{\pi^2}{3}
 \,-\,2x_{\perp j}^2
\right)\, \label{eq:SudakovIntegral}
\end{equation}
where we restored the prefactor from the original matrix-element ratio. 

In this study, we wish to adopt a parameterisation such that the $\alpha_s^AL^\lambda$ coefficients are of order unity. In \eqRef{eq:SudakovIntegral}, the numerical value of the coefficients  are:
\begin{equation}
\alpha_s\left(0.21 \ln^2(x_{\perp j}^2)
\,+\,0.64\ln(x_{\perp j}^2)
\,+\,0.58
\,-\,0.42x_{\perp j}^2\right)~.
\end{equation}
We see that the leading coefficient is actually quite a bit smaller than unity, and there is a relative factor 3 between the leading and the next-to-leading logarithmic term. We can make this more consistent with our assumption by changing to using a linear argument in the logarithms,
\begin{equation}
\frac{C_F \alpha_s}{2\pi}\left( 
 4\ln^2(x_{\perp j})
 \,+\,6\ln(x_{\perp j}) 
 \,+\,6
 \,-\,\frac{\pi^2}{3}
 \,-\,2x_{\perp j}^2
\right)\,, \label{eq:SudakovIntegralLin}
\end{equation}
with numerical coefficients 
\begin{equation}
\alpha_s\left(0.85 \ln^2(x_{\perp j})
\,+\,1.27\ln(x_{\perp j})
\,+\,0.58
\,-\,0.42x_{\perp j}^2\right)\,,
\end{equation}
where both of the logarithmic coefficients are now closer to unity and there is only a relative factor 1.5 between them. 

To estimate the coefficients generated purely by running-coupling effects, we redo the integrals above using a 2-loop running $\alpha_s(p_{\perp j})$. This yields:
\begin{eqnarray}
&& ~~\alpha_s\left(
~~0.85 \ln^2(x_{\perp j})
\,+\,1.27\ln(x_{\perp j})
\right)\nonumber\\
&&+\alpha_s^2 \left(
-1.04 \ln^3(x_{\perp j})
- 1.55 \ln^2(x_{\perp j}) 
-0.70 \ln(x_{\perp j}) 
\right) \nonumber\\
&&+\alpha_s^3 \left(
 ~~1.26 \ln^4(x_{\perp j})
  + 1.48 \ln^3(x_{\perp j})
  + 0.23 \ln^2(x_{\perp j})
 \right) \nonumber\\
 &&+\alpha_s^4 \left(
 -1.54 \ln^5(x_{\perp j})
 - 1.05 \ln^4(x_{\perp j})
 + 0.86 \ln^3(x_{\perp j})
 \right) \nonumber \\
 & & +\ldots 
\end{eqnarray}
We see that these coefficients are all consistent with order 1, although we do note the systematic 20\% growth per order of the LL coefficients, as mentioned in the main body of the paper.

If we also include the so-called CMW factor~\cite{Catani:1990rr} on the first (eikonal) term in the matrix-element ratio,
\begin{equation}
\frac{2}{x_{\perp j}^2} 
~\to~\frac{2}{x_{\perp j}^2}\left( 
1 + \frac{\alpha_s}{2\pi} K_\mrm{CMW}\right)
~=~
~\frac{2}{x_{\perp j}^2}\left( 
1 + \alpha_s \frac{C_A(67-3\pi^2) + 10n_F}{36\pi} \right)
\end{equation}
we get (using $n_F=5$ for all values of $x_{\perp j}$ for simplicity):
\begin{eqnarray}
&& ~~\alpha_s\left(
~~0.85 \ln^2(x_{\perp j})
\,+\,1.27\ln(x_{\perp j})
\right)\nonumber\\
&&+\alpha_s^2 \left(
-1.04 \ln^3(x_{\perp j})
- 1.09 \ln^2(x_{\perp j}) 
-0.70 \ln(x_{\perp j}) 
\right) \nonumber\\
&&+\alpha_s^3 \left(
 ~~1.26 \ln^4(x_{\perp j})
  + 0.91 \ln^3(x_{\perp j})
  + 0.23 \ln^2(x_{\perp j})
 \right) \nonumber\\
 &&+\alpha_s^4 \left(
 -1.54 \ln^5(x_{\perp j})
 - 0.35 \ln^4(x_{\perp j})
 + 0.63 \ln^3(x_{\perp j})
 \right) \nonumber \\
 & & +\ldots ~,
\end{eqnarray}
with coefficients that remain consistent with unity for the orders shown here.

\phantomsection
\addcontentsline{toc}{section}{\sffamily References}
\pagestyle{plain}
\bibliographystyle{JHEP}
\bibliography{bib}

@article{Catani:2001cc,
    author = "Catani, S. and Krauss, F. and Kuhn, R. and Webber, B. R.",
    title = "{QCD matrix elements + parton showers}",
    eprint = "hep-ph/0109231",
    archivePrefix = "arXiv",
    reportNumber = "CERN-TH-2000-367, CAVENDISH-HEP-00-03",
    doi = "10.1088/1126-6708/2001/11/063",
    journal = "JHEP",
    volume = "11",
    pages = "063",
    year = "2001"
}

@article{Nagy:2020rmk,
    author = "Nagy, Zolt{\'a}n and Soper, Davison E.",
    title = "{Summations of large logarithms by parton showers}",
    eprint = "2011.04773",
    archivePrefix = "arXiv",
    primaryClass = "hep-ph",
    reportNumber = "DESY 20-181, DESY-20-181",
    doi = "10.1103/PhysRevD.104.054049",
    journal = "Phys. Rev. D",
    volume = "104",
    number = "5",
    pages = "054049",
    year = "2021"
}

@article{Forshaw:2020wrq,
    author = {Forshaw, Jeffrey R. and Holguin, Jack and Pl{\"a}tzer, Simon},
    title = "{Building a consistent parton shower}",
    eprint = "2003.06400",
    archivePrefix = "arXiv",
    primaryClass = "hep-ph",
    reportNumber = "MAN/HEP/2020/002, UWTHPH-2020-8, MCnet-20-11",
    doi = "10.1007/JHEP09(2020)014",
    journal = "JHEP",
    volume = "09",
    pages = "014",
    year = "2020"
}

@article{Hoche:2017kst,
    author = {H{\"o}che, Stefan and Reichelt, Daniel and Siegert, Frank},
    title = "{Momentum conservation and unitarity in parton showers and NLL resummation}",
    eprint = "1711.03497",
    archivePrefix = "arXiv",
    primaryClass = "hep-ph",
    reportNumber = "SLAC-PUB-17173, MCNET-17-20",
    doi = "10.1007/JHEP01(2018)118",
    journal = "JHEP",
    volume = "01",
    pages = "118",
    year = "2018"
}

@article{Lonnblad:2012ix,
    author = {L{\"o}nnblad, Leif and Prestel, Stefan},
    title = "{Merging Multi-leg NLO Matrix Elements with Parton Showers}",
    eprint = "1211.7278",
    archivePrefix = "arXiv",
    primaryClass = "hep-ph",
    reportNumber = "MCNET-12-17, LU-TP-12-43",
    doi = "10.1007/JHEP03(2013)166",
    journal = "JHEP",
    volume = "03",
    pages = "166",
    year = "2013"
}

@article{Lonnblad:2011xx,
    author = "L{\"o}nnblad, Leif and Prestel, Stefan",
    title = "{Matching Tree-Level Matrix Elements with Interleaved Showers}",
    eprint = "1109.4829",
    archivePrefix = "arXiv",
    primaryClass = "hep-ph",
    doi = "10.1007/JHEP03(2012)019",
    journal = "JHEP",
    volume = "03",
    pages = "019",
    year = "2012"
}

@article{Cooper:2011gk,
    author = "Cooper, B. and Katzy, J. and Mangano, M. L. and Messina, A. and Mijovic, L. and Skands, P.",
    title = "{Importance of a consistent choice of alpha(s) in the matching of AlpGen and Pythia}",
    eprint = "1109.5295",
    archivePrefix = "arXiv",
    primaryClass = "hep-ph",
    reportNumber = "CERN-PH-TH-2011-228, DESY-11-124",
    doi = "10.1140/epjc/s10052-012-2078-y",
    journal = "Eur. Phys. J. C",
    volume = "72",
    pages = "2078",
    year = "2012"
}

@article{Dasgupta:2020fwr,
    author = "Dasgupta, Mrinal and Dreyer, Fr{\'e}d{\'e}ric A. and Hamilton, Keith and Monni, Pier Francesco and Salam, Gavin P. and Soyez, Gregory",
    title = "{Parton showers beyond leading logarithmic accuracy}",
    eprint = "2002.11114",
    archivePrefix = "arXiv",
    primaryClass = "hep-ph",
    reportNumber = "CERN-TH-2020-026",
    doi = "10.1103/PhysRevLett.125.052002",
    journal = "Phys. Rev. Lett.",
    volume = "125",
    number = "5",
    pages = "052002",
    year = "2020"
}

@article{Brooks:2020upa,
    author = "Brooks, Helen and Preuss, Christian T. and Skands, Peter",
    title = "{Sector Showers for Hadron Collisions}",
    eprint = "2003.00702",
    archivePrefix = "arXiv",
    primaryClass = "hep-ph",
    reportNumber = "CoEPP-MN-20-2, MCNET-20-09",
    doi = "10.1007/JHEP07(2020)032",
    journal = "JHEP",
    volume = "07",
    pages = "032",
    year = "2020"
}

@article{Bierlich:2022pfr,
    author = "Bierlich, Christian and others",
    title = "{A comprehensive guide to the physics and usage of PYTHIA 8.3}",
    eprint = "2203.11601",
    archivePrefix = "arXiv",
    primaryClass = "hep-ph",
    reportNumber = "LU-TP 22-16, MCNET-22-04, FERMILAB-PUB-22-227-SCD",
    doi = "10.21468/SciPostPhysCodeb.8",
    journal = "SciPost Phys. Codeb.",
    volume = "2022",
    pages = "8",
    year = "2022"
}

@article{FerrarioRavasio:2023kyg,
    author = "Ferrario Ravasio, Silvia and Hamilton, Keith and Karlberg, Alexander and Salam, Gavin P. and Scyboz, Ludovic and Soyez, Gregory",
    title = "{Parton Showering with Higher Logarithmic Accuracy for Soft Emissions}",
    eprint = "2307.11142",
    archivePrefix = "arXiv",
    primaryClass = "hep-ph",
    reportNumber = "CERN-TH-2023-127, OUTP-23-07P",
    doi = "10.1103/PhysRevLett.131.161906",
    journal = "Phys. Rev. Lett.",
    volume = "131",
    number = "16",
    pages = "161906",
    year = "2023"
}

@article{vanBeekveld:2024wws,
    author = "van Beekveld, Melissa and others",
    title = "{New Standard for the Logarithmic Accuracy of Parton Showers}",
    eprint = "2406.02661",
    archivePrefix = "arXiv",
    primaryClass = "hep-ph",
    reportNumber = "CERN-TH-2024-057, OUTP-24-03P",
    doi = "10.1103/PhysRevLett.134.011901",
    journal = "Phys. Rev. Lett.",
    volume = "134",
    number = "1",
    pages = "011901",
    year = "2025"
}

@article{Ellis:1980wv,
    author = "Ellis, R. Keith and Ross, D. A. and Terrano, A. E.",
    title = "{The Perturbative Calculation of Jet Structure in e+ e- Annihilation}",
    reportNumber = "CALT-68-785",
    doi = "10.1016/0550-3213(81)90165-6",
    journal = "Nucl. Phys. B",
    volume = "178",
    pages = "421--456",
    year = "1981"
}

@article{Hartgring:2013jma,
    author = "Hartgring, L. and Laenen, E. and Skands, P.",
    title = "{Antenna Showers with One-Loop Matrix Elements}",
    eprint = "1303.4974",
    archivePrefix = "arXiv",
    primaryClass = "hep-ph",
    reportNumber = "NIKHEF-2013-004, CERN-PH-TH-2013-038, ITF-UU-13-02",
    doi = "10.1007/JHEP10(2013)127",
    journal = "JHEP",
    volume = "10",
    pages = "127",
    year = "2013"
}

@article{Frixione:2002ik,
    author = "Frixione, Stefano and Webber, Bryan R.",
    title = "{Matching NLO QCD computations and parton shower simulations}",
    eprint = "hep-ph/0204244",
    archivePrefix = "arXiv",
    reportNumber = "CAVENDISH-HEP-02-01, LAPTH-905-02, GEF-TH-2-2002",
    doi = "10.1088/1126-6708/2002/06/029",
    journal = "JHEP",
    volume = "06",
    pages = "029",
    year = "2002"
}

@article{Frederix:2012ps,
    author = "Frederix, Rikkert and Frixione, Stefano",
    title = "{Merging meets matching in MC@NLO}",
    eprint = "1209.6215",
    archivePrefix = "arXiv",
    primaryClass = "hep-ph",
    reportNumber = "CERN-PH-TH-2012-247, ZU-TH-21-12",
    doi = "10.1007/JHEP12(2012)061",
    journal = "JHEP",
    volume = "12",
    pages = "061",
    year = "2012"
}

@article{Frixione:2007vw,
    author = "Frixione, Stefano and Nason, Paolo and Oleari, Carlo",
    title = "{Matching NLO QCD computations with Parton Shower simulations: the POWHEG method}",
    eprint = "0709.2092",
    archivePrefix = "arXiv",
    primaryClass = "hep-ph",
    reportNumber = "BICOCCA-FT-07-9, GEF-TH-21-2007",
    doi = "10.1088/1126-6708/2007/11/070",
    journal = "JHEP",
    volume = "11",
    pages = "070",
    year = "2007"
}

@article{vanBeekveld:2025lpz,
    author = "van Beekveld, Melissa and Ferrario Ravasio, Silvia and Helliwell, Jack and Karlberg, Alexander and Salam, Gavin P. and Scyboz, Ludovic and Soto-Ontoso, Alba and Soyez, Gregory and Zanoli, Silvia",
    title = "{Logarithmically-accurate and positive-definite NLO shower matching}",
    eprint = "2504.05377",
    archivePrefix = "arXiv",
    primaryClass = "hep-ph",
    reportNumber = "CERN-TH-2025-004, OUTP-25-01P, Nikhef 2025-003",
    doi = "10.1007/JHEP10(2025)038",
    journal = "JHEP",
    volume = "10",
    pages = "038",
    year = "2025"
}

@article{Jadach:2015mza,
    author = "Jadach, S. and P{\l}aczek, W. and Sapeta, S. and Si{\'o}dmok, A. and Skrzypek, M.",
    title = "{Matching NLO QCD with parton shower in Monte Carlo scheme {\textemdash} the KrkNLO method}",
    eprint = "1503.06849",
    archivePrefix = "arXiv",
    primaryClass = "hep-ph",
    reportNumber = "IFJPAN-IV-2015-1, CERN-PH-TH-2015-061, MCNET-15-06",
    doi = "10.1007/JHEP10(2015)052",
    journal = "JHEP",
    volume = "10",
    pages = "052",
    year = "2015"
}

@article{Lonnblad:2012ng,
    author = "L{\"o}nnblad, Leif and Prestel, Stefan",
    title = "{Unitarising Matrix Element + Parton Shower merging}",
    eprint = "1211.4827",
    archivePrefix = "arXiv",
    primaryClass = "hep-ph",
    reportNumber = "LU-TP-12-42, MCNET-12-14",
    doi = "10.1007/JHEP02(2013)094",
    journal = "JHEP",
    volume = "02",
    pages = "094",
    year = "2013"
}

@article{Gehrmann-DeRidder:2005btv,
    author = "Gehrmann-De Ridder, A. and Gehrmann, T. and Glover, E. W. Nigel",
    title = "{Antenna subtraction at NNLO}",
    eprint = "hep-ph/0505111",
    archivePrefix = "arXiv",
    reportNumber = "ZU-TH-07-05, IPPP-05-18",
    doi = "10.1088/1126-6708/2005/09/056",
    journal = "JHEP",
    volume = "09",
    pages = "056",
    year = "2005"
}

@article{Dinsdale:2007mf,
    author = "Dinsdale, Michael and Ternick, Marko and Weinzierl, Stefan",
    title = "{Parton showers from the dipole formalism}",
    eprint = "0709.1026",
    archivePrefix = "arXiv",
    primaryClass = "hep-ph",
    reportNumber = "MZ-TH-07-14",
    doi = "10.1103/PhysRevD.76.094003",
    journal = "Phys. Rev. D",
    volume = "76",
    pages = "094003",
    year = "2007"
}

@article{Nagy:2007ty,
    author = "Nagy, Zoltan and Soper, Davison E.",
    title = "{Parton showers with quantum interference}",
    eprint = "0706.0017",
    archivePrefix = "arXiv",
    primaryClass = "hep-ph",
    reportNumber = "CERN-PH-TH-2007-082",
    doi = "10.1088/1126-6708/2007/09/114",
    journal = "JHEP",
    volume = "09",
    pages = "114",
    year = "2007"
}

@article{Marchesini:1983bm,
    author = "Marchesini, G. and Webber, B. R.",
    title = "{Simulation of QCD Jets Including Soft Gluon Interference}",
    reportNumber = "CERN-TH-3525",
    doi = "10.1016/0550-3213(84)90463-2",
    journal = "Nucl. Phys. B",
    volume = "238",
    pages = "1--29",
    year = "1984"
}

@article{Gieseke:2003rz,
    author = "Gieseke, Stefan and Stephens, P. and Webber, Bryan",
    title = "{New formalism for QCD parton showers}",
    eprint = "hep-ph/0310083",
    archivePrefix = "arXiv",
    reportNumber = "CAVENDISH-HEP-03-18, CERN-TH-2003-239",
    doi = "10.1088/1126-6708/2003/12/045",
    journal = "JHEP",
    volume = "12",
    pages = "045",
    year = "2003"
}

@article{Platzer:2009jq,
    author = "Platzer, Simon and Gieseke, Stefan",
    title = "{Coherent Parton Showers with Local Recoils}",
    eprint = "0909.5593",
    archivePrefix = "arXiv",
    primaryClass = "hep-ph",
    reportNumber = "HERWIG-09-06, KA-TP-15-2009, MCNET-09-15",
    doi = "10.1007/JHEP01(2011)024",
    journal = "JHEP",
    volume = "01",
    pages = "024",
    year = "2011"
}

@article{Alioli:2012fc,
    author = "Alioli, Simone and Bauer, Christian W. and Berggren, Calvin J. and Hornig, Andrew and Tackmann, Frank J. and Vermilion, Christopher K. and Walsh, Jonathan R. and Zuberi, Saba",
    title = "{Combining Higher-Order Resummation with Multiple NLO Calculations and Parton Showers in GENEVA}",
    eprint = "1211.7049",
    archivePrefix = "arXiv",
    primaryClass = "hep-ph",
    reportNumber = "DESY-12-221",
    doi = "10.1007/JHEP09(2013)120",
    journal = "JHEP",
    volume = "09",
    pages = "120",
    year = "2013"
}

@article{Hoche:2014uhw,
    author = {H{\"o}che, Stefan and Li, Ye and Prestel, Stefan},
    title = "{Drell-Yan lepton pair production at NNLO QCD with parton showers}",
    eprint = "1405.3607",
    archivePrefix = "arXiv",
    primaryClass = "hep-ph",
    reportNumber = "SLAC-PUB-15961, DESY-14-073, MCNET-14-10",
    doi = "10.1103/PhysRevD.91.074015",
    journal = "Phys. Rev. D",
    volume = "91",
    number = "7",
    pages = "074015",
    year = "2015"
}

@article{Lopez-Villarejo:2011pwr,
    author = "Lopez-Villarejo, J. J. and Skands, Peter Z.",
    title = "{Efficient Matrix-Element Matching with Sector Showers}",
    eprint = "1109.3608",
    archivePrefix = "arXiv",
    primaryClass = "hep-ph",
    reportNumber = "CERN-TH-2011-202, MCNET-11-22",
    doi = "10.1007/JHEP11(2011)150",
    journal = "JHEP",
    volume = "11",
    pages = "150",
    year = "2011"
}

@article{Giele:2007di,
    author = "Giele, Walter T. and Kosower, David A. and Skands, Peter Z.",
    title = "{A simple shower and matching algorithm}",
    eprint = "0707.3652",
    archivePrefix = "arXiv",
    primaryClass = "hep-ph",
    reportNumber = "FERMILAB-PUB-07-160-T, SACLAY-SPHT-T07-107",
    doi = "10.1103/PhysRevD.78.014026",
    journal = "Phys. Rev. D",
    volume = "78",
    pages = "014026",
    year = "2008"
}

@article{Hoche:2015sya,
    author = {H{\"o}che, Stefan and Prestel, Stefan},
    title = "{The midpoint between dipole and parton showers}",
    eprint = "1506.05057",
    archivePrefix = "arXiv",
    primaryClass = "hep-ph",
    reportNumber = "SLAC-PUB-16304, MCNET-15-13",
    doi = "10.1140/epjc/s10052-015-3684-2",
    journal = "Eur. Phys. J. C",
    volume = "75",
    number = "9",
    pages = "461",
    year = "2015"
}

@article{Monni:2019whf,
    author = "Monni, Pier Francesco and Nason, Paolo and Re, Emanuele and Wiesemann, Marius and Zanderighi, Giulia",
    title = "{MiNNLO$_{PS}$: a new method to match NNLO QCD to parton showers}",
    eprint = "1908.06987",
    archivePrefix = "arXiv",
    primaryClass = "hep-ph",
    reportNumber = "CERN-TH-2019-117, LAPTH-042/19, MPP-2019-177",
    doi = "10.1007/JHEP05(2020)143",
    journal = "JHEP",
    volume = "05",
    pages = "143",
    year = "2020",
    note = "[Erratum: JHEP 02, 031 (2022)]"
}

@article{Mangano:2006rw,
    author = "Mangano, Michelangelo L. and Moretti, Mauro and Piccinini, Fulvio and Treccani, Michele",
    title = "{Matching matrix elements and shower evolution for top-quark production in hadronic collisions}",
    eprint = "hep-ph/0611129",
    archivePrefix = "arXiv",
    reportNumber = "CERN-PH-TH-2006-232, FNT-T-2006-09",
    doi = "10.1088/1126-6708/2007/01/013",
    journal = "JHEP",
    volume = "01",
    pages = "013",
    year = "2007"
}

@article{Hamilton:2012rf,
    author = "Hamilton, Keith and Nason, Paolo and Oleari, Carlo and Zanderighi, Giulia",
    title = "{Merging H/W/Z + 0 and 1 jet at NLO with no merging scale: a path to parton shower + NNLO matching}",
    eprint = "1212.4504",
    archivePrefix = "arXiv",
    primaryClass = "hep-ph",
    reportNumber = "CERN-PH-TH-2012-356",
    doi = "10.1007/JHEP05(2013)082",
    journal = "JHEP",
    volume = "05",
    pages = "082",
    year = "2013"
}

@article{Lonnblad:2001iq,
    author = "L{\"o}nnblad, Leif",
    title = "{Correcting the color dipole cascade model with fixed order matrix elements}",
    eprint = "hep-ph/0112284",
    archivePrefix = "arXiv",
    reportNumber = "LU-TP-01-38",
    doi = "10.1088/1126-6708/2002/05/046",
    journal = "JHEP",
    volume = "05",
    pages = "046",
    year = "2002"
}

@article{Lonnblad:1992tz,
    author = "L{\"o}nnblad, Leif",
    title = "{ARIADNE version 4: A Program for simulation of QCD cascades implementing the color dipole model}",
    reportNumber = "DESY-92-046",
    doi = "10.1016/0010-4655(92)90068-A",
    journal = "Comput. Phys. Commun.",
    volume = "71",
    pages = "15--31",
    year = "1992"
}

@article{Gustafson:1987rq,
    author = "Gustafson, Gosta and Pettersson, Ulf",
    title = "{Dipole Formulation of QCD Cascades}",
    reportNumber = "LU-TP-87-9",
    doi = "10.1016/0550-3213(88)90441-5",
    journal = "Nucl. Phys. B",
    volume = "306",
    pages = "746--758",
    year = "1988"
}

@article{Banfi:2004yd,
    author = "Banfi, Andrea and Salam, Gavin P. and Zanderighi, Giulia",
    title = "{Principles of general final-state resummation and automated implementation}",
    eprint = "hep-ph/0407286",
    archivePrefix = "arXiv",
    reportNumber = "FERMILAB-PUB-04-116-T, LPTHE-04-16, NIKHEF-2004-005",
    doi = "10.1088/1126-6708/2005/03/073",
    journal = "JHEP",
    volume = "03",
    pages = "073",
    year = "2005"
}

@article{vanBeekveld:2026uxl,
    author = {van Beekveld, Melissa and Bothmann, Enrico and Buckley, Andy and G{\"u}tschow, Christian and Skands, Peter and Winterhalder, Ramon},
    title = "{The Monte Carlo Ecosystem in High-Energy Physics: A Primer}",
    eprint = "2605.16036",
    archivePrefix = "arXiv",
    primaryClass = "hep-ph",
    reportNumber = "MCNET-26-11, TIF-UNIMI-2026-7",
    month = "5",
    year = "2026"
}

@article{Dasgupta:2018nvj,
    author = "Dasgupta, Mrinal and Dreyer, Fr{\'e}d{\'e}ric A. and Hamilton, Keith and Monni, Pier Francesco and Salam, Gavin P.",
    title = "{Logarithmic accuracy of parton showers: a fixed-order study}",
    eprint = "1805.09327",
    archivePrefix = "arXiv",
    primaryClass = "hep-ph",
    reportNumber = "CERN-TH-2018-113",
    doi = "10.1007/JHEP09(2018)033",
    journal = "JHEP",
    volume = "09",
    pages = "033",
    year = "2018",
    note = "[Erratum: JHEP 03, 083 (2020)]"
}

@article{Altmann:2025yip,
    author = "Altmann, Javira and Li, Hai Tao and Scyboz, Ludovic and Skands, Peter",
    title = "{Sudakov evolution without unitarity}",
    eprint = "2507.00111",
    archivePrefix = "arXiv",
    primaryClass = "hep-ph",
    doi = "10.1140/epjc/s10052-025-14576-1",
    journal = "Eur. Phys. J. C",
    volume = "85",
    number = "8",
    pages = "840",
    year = "2025"
}

@article{Catani:1990rr,
    author = "Catani, S. and Webber, B. R. and Marchesini, G.",
    title = "{QCD coherent branching and semiinclusive processes at large x}",
    reportNumber = "CAVENDISH-HEP-90-11, UPRF-90-280",
    doi = "10.1016/0550-3213(91)90390-J",
    journal = "Nucl. Phys. B",
    volume = "349",
    pages = "635--654",
    year = "1991"
}

@article{Sjostrand:2004ef,
    author = "Sj{\"o}strand, T. and Skands, Peter Z.",
    title = "{Transverse-momentum-ordered showers and interleaved multiple interactions}",
    eprint = "hep-ph/0408302",
    archivePrefix = "arXiv",
    reportNumber = "LU-TP-04-29",
    doi = "10.1140/epjc/s2004-02084-y",
    journal = "Eur. Phys. J. C",
    volume = "39",
    pages = "129--154",
    year = "2005"
}

@article{Helliwell:2025krf,
    author = "Helliwell, Jack and Scyboz, Ludovic and Skands, Peter",
    title = "{Timelike showers with jet recoils}",
    eprint = "2512.07370",
    archivePrefix = "arXiv",
    primaryClass = "hep-ph",
    doi = "10.1007/JHEP05(2026)284",
    journal = "JHEP",
    volume = "05",
    pages = "284",
    year = "2026"
}

@article{Preuss:2024vyu,
    author = "Preuss, Christian T.",
    title = "{A partitioned dipole-antenna shower with improved transverse recoil}",
    eprint = "2403.19452",
    archivePrefix = "arXiv",
    primaryClass = "hep-ph",
    doi = "10.1007/JHEP07(2024)161",
    journal = "JHEP",
    volume = "07",
    pages = "161",
    year = "2024"
}

@article{Herren:2022jej,
    author = {Herren, Florian and H{\"o}che, Stefan and Krauss, Frank and Reichelt, Daniel and Schoenherr, Marek},
    title = "{A new approach to color-coherent parton evolution}",
    eprint = "2208.06057",
    archivePrefix = "arXiv",
    primaryClass = "hep-ph",
    reportNumber = "FERMILAB-PUB-22-556-T, IPPP/22/58, MCNET-22-14",
    doi = "10.1007/JHEP10(2023)091",
    journal = "JHEP",
    volume = "10",
    pages = "091",
    year = "2023"
}

@article{Hamilton:2020rcu,
    author = "Hamilton, Keith and Medves, Rok and Salam, Gavin P. and Scyboz, Ludovic and Soyez, Gregory",
    title = "{Colour and logarithmic accuracy in final-state parton showers}",
    eprint = "2011.10054",
    archivePrefix = "arXiv",
    primaryClass = "hep-ph",
    doi = "10.1007/JHEP03(2021)041",
    journal = "JHEP",
    volume = "03",
    number = "041",
    pages = "041",
    year = "2021"
}

\end{document}